\documentclass[pre,twocolumn,aps,amsmath,amssymb,superscriptaddress]{revtex4-1}

\usepackage[hypertexnames=true,colorlinks]{hyperref}
\hypersetup{pdffitwindow=true,
             linkcolor=blue,
             citecolor=blue,
             urlcolor=blue}
\usepackage{times}
\usepackage{mathptmx}
\usepackage{miller}
\usepackage{graphics}
\usepackage{graphicx}
\usepackage{nicefrac}
\usepackage{amssymb}
\usepackage{xcolor}
\usepackage{wasysym}
\definecolor{bgreen}{rgb}{0.6, 0.4, 0.8}

\newcommand{\etal}{\textit{et al.}}
\newcommand{\fref}[1]{Fig.~\ref{#1}}

\newcommand{\tref}[1]{Table~\ref{#1}}
\newcommand{\Tref}[1]{Table~\ref{#1}}

\newcommand{\eref}[1]{Eq.~(\ref{#1})}

\newcommand{\yes}{\CIRCLE}
\newcommand{\yesno}{\RIGHTcircle}
\newcommand{\no}{\Circle}
\newcommand{\ocite}[1]{\cite{#1}}

\begin{document}

\title{\{110\} planar faults in strained bcc metals: Origins and implications \\ of a commonly observed artifact of classical potentials}

\author{Johannes J. M\"oller}
\email[Corresponding author: ]{johannes.moeller@iwm.fraunhofer.de}
\affiliation{Friedrich-Alexander-Universit\"at Erlangen-N\"urnberg, Department of Materials Science and Engineering, Institute I, Martensstr. 5, D-91058 Erlangen, Germany}
\affiliation{Fraunhofer Institute for Mechanics of Materials IWM, W\"ohlerstr. 11, D-79108 Freiburg, Germany}

\author{Matous Mrovec}
\affiliation{Interdisciplinary Centre for Advanced Materials Simulation, Ruhr-University Bochum, Universit\"atsstr. 150, D-44780 Bochum, Germany}

\author{Ivan Bleskov} 
\author{J\"org Neugebauer}
\affiliation{Computational Materials Design, Max-Planck-Institut f\"ur Eisenforschung GmbH, Max-Planck-Str. 1, D-40237 D\"usseldorf, Germany}

\author{Thomas Hammerschmidt}
\author{Ralf Drautz}
\affiliation{Interdisciplinary Centre for Advanced Materials Simulation, Ruhr-University Bochum, Universit\"atsstr. 150, D-44780 Bochum, Germany}

\author{Christian Els\"asser} 
\affiliation{Fraunhofer Institute for Mechanics of Materials IWM, W\"ohlerstr. 11, D-79108 Freiburg, Germany}
\affiliation{University of Freiburg, Freiburg Materials Research Center, Stefan-Meier-Str. 21, D-79104 Freiburg, Germany}

\author{Tilmann Hickel} 
\affiliation{Computational Materials Design, Max-Planck-Institut f\"ur Eisenforschung GmbH, Max-Planck-Str. 1, D-40237 D\"usseldorf, Germany}

\author{Erik Bitzek}
\affiliation{Friedrich-Alexander-Universit\"at Erlangen-N\"urnberg, Department of Materials Science and Engineering, Institute I, Martensstr. 5, D-91058 Erlangen, Germany}

\begin{abstract}
Large-scale atomistic simulations with classical potentials can provide valuable insights into microscopic deformation mechanisms and defect-defect interactions in materials. 
Unfortunately, these assets often come with the uncertainty of whether the observed mechanisms are based on realistic physical phenomena or whether they are artifacts of the employed material models. 
One such example is the often reported occurrence of stable planar faults (PFs) in body-centered cubic (bcc) metals subjected to high strains, e.g., at crack tips or in strained nano-objects.
In this paper, we study the strain dependence of the generalized stacking fault energy (GSFE) of \hkl{110} planes in various bcc metals with material models of increasing sophistication, i.e., (modified) embedded atom method, angular-dependent, Tersoff, and bond-order potentials as well as density functional theory.
We show that under applied tensile strains the GSFE curves of many classical potentials exhibit a local minimum which gives rise to the formation of stable PFs.
These PFs do not appear when more sophisticated material models are used and have thus to be regarded as artifacts of the potentials. We demonstrate that the local GSFE minimum is not formed for reasons of symmetry and we recommend including the determination of the strain-dependent \hkl(110) GSFE as a benchmark for newly developed potentials. 
\end{abstract}

\maketitle
\section{Introduction}


Atomic-scale modeling of materials can provide fundamental information about microscopic deformation mechanisms and defect-defect interactions \cite{Mis10},
which can ultimately stimulate the development of higher-scale models with input parameters from atomistic simulations \cite{Nee01,Leyson2010}.
At the same time, results of atomic-level simulations are often subject to some uncertainty, particularly in case of classical molecular dynamics (MD) simulations.
This uncertainty mainly arises from the question whether the interatomic potential employed in the simulations correctly reflects the material's response to the applied loads.
However, studies of plastic deformation and fracture necessitate the simulation of many millions of atoms \cite{Has03,Bit09,Hor10,Bit13},
where classical interatomic models are without alternatives.
To check if the performed simulations are independent of the used material model
is therefore of prior importance for the transferability and acceptance of the obtained results.
It is important to note that the common approach to only compare different potentials of the same type 
is generally not sufficient as all potentials of the same type might suffer from the same problems. 

In this paper, we use various well-established material models to determine the strain-dependent 
generalized stacking fault energy (GSFE) curves for body-centered cubic (bcc) transition metals. 
The GSFE (also known as "$\gamma$ surface") \cite{Vit68} is determined by rigidly shifting two half crystals with respect to each other, 
typically under stress-free boundary conditions.
Local GSFE minima and maxima indicate stable (and unstable) stacking faults (SSFs), respectively.
Whether a material has a local GSFE minimum can be theoretically inferred from its crystal symmetry. 
This principle, which was originally established by Neumann \cite{Neumann1885}, states that GSFE minima are only possible where two non-parallel symmetry planes intersect the crystal's glide plane.
Unlike bcc materials, face-centered cubic (fcc) crystals contain such crossings of symmetry planes in their slip planes and therefore exhibit SSFs. 
Despite the idealized nature of the GSFE, the presence of SSFs, their energies, and relation to the unstable stacking fault energy turned out to be a good indicator for the nature of the material's slip behavior \cite{Vit70,Christian2002a,Tschopp2007}.  
This is especially evident for materials with local GSFE minima, such as the fcc metals Al, Cu, and Ni, 
for which Brandl \etal{} \cite{Brandl2007} have already determined the strain-dependent GSFEs for isotropic and simple shear strains. 
The strain dependence of the GSFE in bcc crystals, however, is still unknown.   

For unstrained bcc crystals, it has been known for a long time that the formation of SSFs by dislocation dissociation is not possible \cite{Vit68}.
However, for situations where the elastic stress state deviates strongly from stress-free conditions
this common conception needs not to be valid.
In such extreme scenarios, e.g., at crack tips or in nanostructures, 
the following dislocation dissociation might therefore be possible \cite{Kro64}:
\begin{eqnarray}
\frac{a_0}{2}\text{\hkl[1-11]} & \rightarrow & \frac{a_0}{8}\text{\hkl[1-10]} + \frac{a_0}{4}\text{\hkl[1-12]} + \frac{a_0}{8}\text{\hkl[1-10]}
\label{eq:dissociation_bcc}
\end{eqnarray}
This possibility has been theoretically proposed and discussed by a number of research groups in the early 1960s \cite{Cru61,Cru62,Coh62,Kro64,Kro67,Bog64} based on hard-sphere models.
Additionally, some experimental studies indeed gave an indication for the existence of dissociated dislocations on \hkl{110} planes in Fe \cite{Smi69} and Nb \cite{Rot71}. 

\begin{figure}
\includegraphics[width=0.5\textwidth]{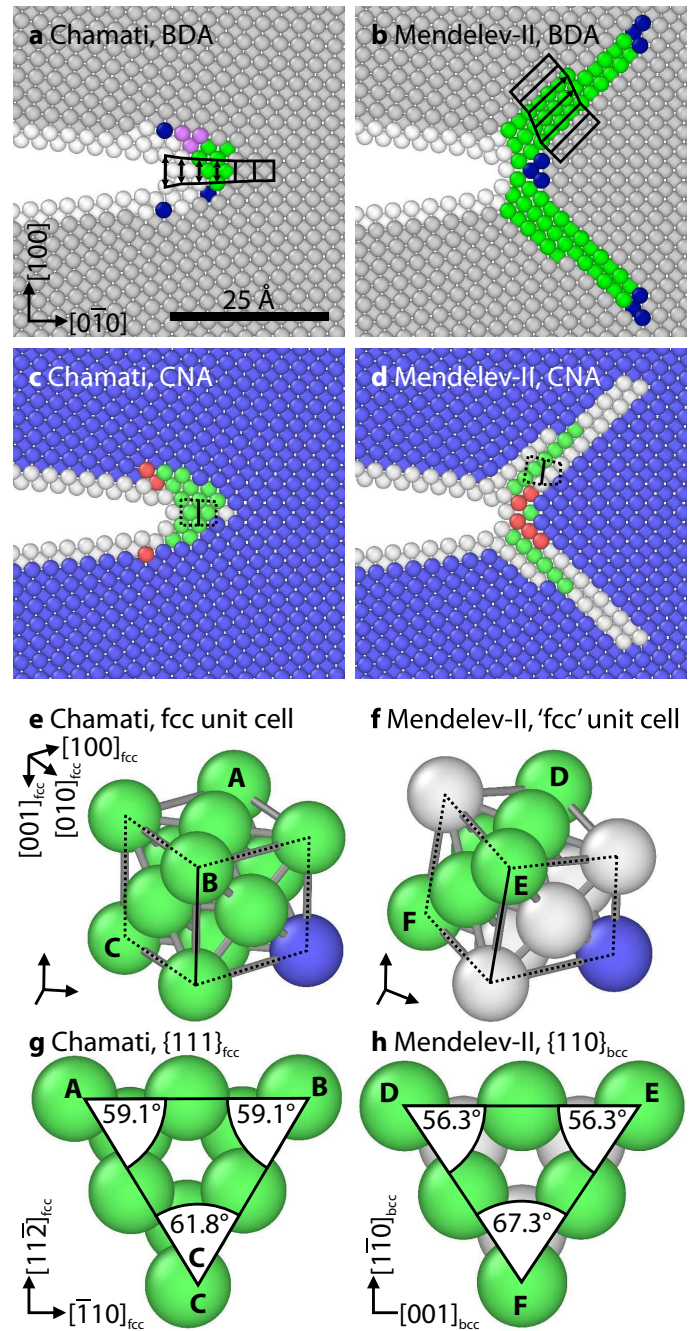}
\caption{Comparison of atomistic configurations in Fe modeled with the Chamati (left, no PF formation) and Mendelev-II potentials (right, PF formation), 
see Ref.~\cite{Moe14msmse} for details.
(a) and (b), \hkl(100)\hkl[001] crack system directly before brittle fracture, in BDA \cite{Moe2016,bda} coloring (grey: bulk; white: surface; green: \hkl{110} PFs; blue: non-screw dislocation), the main deformation paths are marked with black lines;
(c) and (d), same configurations in CNA \cite{Hon87} coloring (blue: bcc; green: fcc; red: hcp; white: other);
(e) and (f), possible fcc unit cells extracted from the marked regions in (c) and (d);
(g) and (h), top views on possible \hkl{111}$_\text{fcc}$ planes extracted from the unit cells in (e) and (f);
the characteristic angles between possible \hkl<110>$_\text{fcc}$ bonds deviate less than 2$^\circ$ from the ideal fcc angles for the Chamati potential
($\sphericalangle\text{ABC}$~=~$\sphericalangle\text{BCA}$~=~$\sphericalangle\text{CAB}$~=~60$^\circ$);
for the Mendelev-II potential, the characteristic angles are still closer to the ideal bcc values ($\sphericalangle\text{DEF}$~=~$\sphericalangle\text{FDE}$~$\approx$~54.7$^\circ$, $\sphericalangle\text{EFD}$~$\approx$~70.5$^\circ$), where the maximum deviation is less than 4$^\circ$ 
(maximum deviation from the characteristic fcc angles $>7^\circ$).
}
\label{fig:cracktips_fcc-vs-110pf}
\end{figure}

More than two decades after these experimental works, the formation of PFs has now been observed in atomistic simulations 
of highly strained systems, e.g., at straight crack tips \cite{Lat03,Bor11,Vat11,Vatne2011,Moe14msmse,Motasem2016}, 
highly curved, penny-shaped crack fronts \cite{Ers12,Moe15}, and nanobeams subjected to bending \cite{Yang2016} and tension \cite{Zhang2012a} as well as containing notches \cite{Skogsrud2015}.
The formation of PFs at straight crack tips is  exemplified in Figs. \ref{fig:cracktips_fcc-vs-110pf}(a) and \ref{fig:cracktips_fcc-vs-110pf}(b) for two different embedded atom method (EAM) potentials for Fe. 
Whereas for the Chamati potential \cite{Cham06}, \fref{fig:cracktips_fcc-vs-110pf}(a),
no extended planar defects appear on \hkl{110} planes (inclined by $\pm$45$^\circ$), 
these defects are clearly visible (green colored atoms) for 
the Mendelev-II potential \cite{Men03}; \fref{fig:cracktips_fcc-vs-110pf}(b).
Upon further loading, the cracks modeled with the Chamati potential propagate on the original \hkl(100) crack plane whereas with the Mendelev-II potential the crack kinks onto one of the inclined \hkl{110} planes \cite{Moe14msmse}.
Examples for the formation of PFs at curved crack fronts and in bent nanowires are provided in the [Supplemental Material].
An earlier work of some of the authors \cite{Moe14msmse} has shown that for many semi-empirical interatomic potentials of the EAM type \cite{Daw84,Daw89}, the formation of these unexpected PFs can be traced back to the strain-induced formation of a local GSFE minimum at relative displacements of approximately $a_0/6$\hkl[1-10], 
which is in the range of the proposed displacement vector in \eref{eq:dissociation_bcc}. 

The atomistic configurations in Figs. \ref{fig:cracktips_fcc-vs-110pf}(a) and  \ref{fig:cracktips_fcc-vs-110pf}(b)  are analyzed using the bcc defect analysis (BDA) \cite{Moe2016}
that was specifically developed to distinguish common defects in bcc structures. 
In contrast, the well-known common neighbor analysis (CNA) \cite{Hon87}, which was used in many of the references above \cite{Lat03,Bor11,Vat11,Vatne2011,Moe14msmse,Ers12,Moe15,Skogsrud2015,Zhang2012a,Yang2016,Motasem2016}, 
identifies the defective regions as belonging to the fcc crystal structure, cf. Figs. \ref{fig:cracktips_fcc-vs-110pf}(c) and  \ref{fig:cracktips_fcc-vs-110pf}(d).
A detailed analysis of the defect structures, however, reveals that the \hkl{110} habit plane of the defects is still closer to \hkl{110}$_\text{bcc}$ planes than to \hkl{111}$_\text{fcc}$ planes.
This can clearly be seen by directly comparing the defective regions for the two potentials in Figs. \ref{fig:cracktips_fcc-vs-110pf}(e)--\ref{fig:cracktips_fcc-vs-110pf}(h).
The misclassification as fcc explains why PFs were often not labeled as such and instead discussed in the context of bcc $\rightarrow$ fcc transformations \cite{Lat03,Vat11,Bor11,Motasem2016}, such as the inverse Bain path \cite{Bain24}.
We will therefore refer to this defect type only as "planar faults" (PF) in the following.
The reader should, however, keep in mind that the same type of planar defects might be labeled as "fcc formation elsewhere.

The objective of this paper is to clarify if there is a physical reason for the strain-dependent formation of PFs in bcc metals. 
For this purpose, we determine the \hkl(110) GSFE in the \hkl[1-10] and \hkl[1-11] directions (so-called "$\gamma$ lines" or "1D GSFE profiles") for systematically varied strain states using different EAM-type potentials as well as several atomistic material models with increasingly sophisticated descriptions of the bonding situations: Modified EAM (MEAM) potentials, angular-dependent potentials (ADPs), Tersoff potentials, bond-order potentials (BOPs), and density functional theory (DFT). 
\tref{tab:summary_models} summarizes the contributions for modeling atomic interactions that are explicitly included or not taken into account 
in the different material models.
Knowledge about the underlying physical reason for the formation of bcc PFs 
is of fundamental importance for the reliability of the results and conclusions in Refs.~ 
\cite{Lat03,Bor11,Vat11,Vatne2011,Moe14msmse,Ers12,Moe15,Skogsrud2015,Zhang2012a,Yang2016}, their transferability to higher scales \cite{Gilbert2011,Cof08b,Tahir2013}, and the development of new interatomic potentials for bcc metals \cite{Ercolessi1994}.

The paper is organized as follows: 
in Sec. \ref{sec:methods}, we introduce the various computational methods for determining $\gamma$ surfaces and $\gamma$ lines. 
The simulation results are presented in Sec. \ref{sec:results_discussion}.
The discussion in Sec. \ref{sec:discussion} focuses on possible explanations for the strain-dependent formation of a local GSFE minimum for EAM-type potentials and how this might be prevented during potential fitting.
Finally, a brief summary of the paper is given in Sec. \ref{sec:summary}.

\begin{table}
\caption{Summary of contributions for modeling atomic interactions explicitly included (\yes) or not taken into account (\no)
in the different material models. 
\label{tab:summary_models}}
\begin{tabular}{cccccc}
\hline
\hline
Model  & 	Radial & Directional	& Bond order	&Magnetism	& Electronic  \\
 & 	distance & bonding	& &	& structure \\
\hline
EAM/FS	& \yes	& \no	& \no	& \no	& \no \\
MEAM		& \yes	& \yes	& \no	& \no	& \no \\
ADP		& \yes	& \yes	& \no	& \no	& \no \\
Tersoff	& \yes	& \yes	& \yes	& \no	& \no \\
BOP		& \yes	& \yes	& \yes	& \yes	& \yesno\footnote{Basic electronic structure captured} \\
DFT		& \yes	& \yes	& \yes	& \yes	& \yes \\
\hline
\hline
\end{tabular}
\end{table}

\section{Methods}
\label{sec:methods}

The 1D \hkl{110} GSFE profiles ($\gamma$ lines) along \hkl[1-10] and \hkl[1-11] directions
were determined as functions of different uniaxial and equi-biaxial strains in various bcc metals.
The setup geometries are described in Sec.~\ref{sec:setup}. 
For Fe, different material models with increasing complexity were used: EAM, MEAM, ADP, Tersoff, BOP, and DFT. 
For the other bcc metals, EAM potentials were compared to DFT. 
The main concepts of the different models are described in Secs.~\ref{sec:EAM}--\ref{sec:DFT}.
\tref{tab:summary_models} provides a comprehensive summary of the different contributions for modeling atomic interactions 
that are explicitly included in the models.

\subsection{Setup geometries}
\label{sec:setup}

Uniaxial strains (US) are applied normal to the PF plane in the range $\varepsilon_{[110]}=0.0$--$10.0~\%$ (tension) with strain increments of $\Delta\varepsilon=1.0~\%$.
Equi-biaxial strains (EBS) contain an equally high additional strain component in \hkl[1-10] direction, i.e., $\varepsilon_{[1\bar{1}0]}=\varepsilon_{[110]}$.
Both strain ranges and directions are chosen to model the high strains that are present at a crack tip; see e.g., \fref{fig:cracktips_fcc-vs-110pf}.

\begin{figure}
\includegraphics[width=0.5\textwidth]{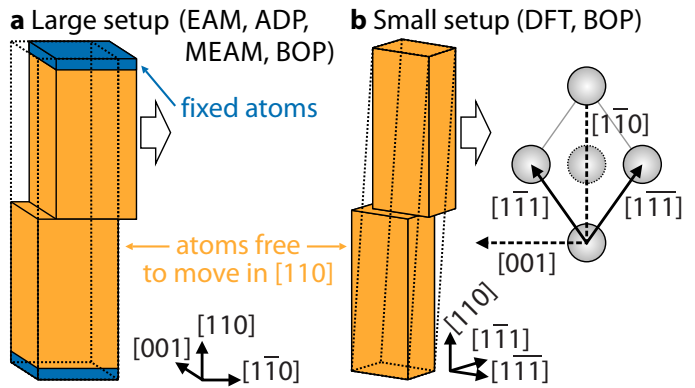}
\caption{Simulation setups for determining $\gamma$ surfaces and $\gamma$ lines. 
The large setup (a) is periodic in \hkl[1-10] and \hkl[001] directions with fixed regions at the top and bottom in the \hkl[110] direction;
the box size is larger than 2$r_\text{cut}$ ($r_\text{cut}$, cutoff radius of the respective potential) in periodic directions and 
approximately 60 unit cells along the long axis;
typical simulation boxes consist of hundreds to thousands of atoms and are used in simulations with classical interatomic potentials.
The small setup \cite{Paxton1992,All96} (b) is periodic in all box directions (dotted lines);
the box size is each one \hkl<111> periodicity distance along the two short box axes and eight periodicity distances along the long \hkl[110] direction;
the box contains usually tens of atoms and is suited for DFT calculations (and simulations using BOPs for comparison);
by adjusting the direction of the long box vector in each shearing step, only one stacking fault is formed in the plane where the two rigid blocks are sheared against each other.
In both setups, atomic motion during the relaxation process is limited to directions perpendicular to the \hkl(110) plane.
}
\label{fig:setups} 
\end{figure}

Figure \ref{fig:setups} displays two different setup geometries for determining the $\gamma$ surfaces and the $\gamma$ lines.
The "large" setup in \fref{fig:setups}(a) is primarily suited for simulations using classical potentials (Secs.~\ref{sec:EAM}--\ref{sec:Tersoff}).
It contains hundreds to thousands of atoms and fulfills the minimum image convention, namely that the box sizes within the PF plane are larger than 2$r_\text{cut}$. 
Along these lateral box axes, periodic boundary conditions (PBC) are used.
The boundary conditions in the direction perpendicular to the PF plane are non-periodic,
which leads to two free surfaces at the top and bottom of the configuration.
During energy minimization, the lateral box sizes are kept fixed, meaning that Poisson contraction is not allowed. 
Most simulation boxes are constructed such that they contain $3 \times 60 \times 4$ unit cells in the \hkl[1-10], \hkl[110], and \hkl[001] directions, respectively.

The "small" setup in \fref{fig:setups}(b) is mainly intended for the more sophisticated descriptions of atomic bonding (Secs.~\ref{sec:BOP} and \ref{sec:DFT}).
This setup is periodic in all directions and has box vectors in the \hkl[1-11] and \hkl[1-1-1] directions.
The long box vector, which is initially normal to the SF plane, is adjusted at each displacement step in such a way that the crystal structure is coherent at both ends of the simulation box.
In other words, the simulation box contains only the stacking fault in the center, but no additional surface or interface at the top or bottom.
The majority of the simulation boxes constructed in this manner contains $1 \times 8 \times 1$ unit cells in the \hkl[1-11], \hkl[110], and \hkl[1-1-1] directions, respectively.
Further details on this simulation setup can be found in Refs.~\cite{Paxton1992,All96}.

To determine the $\gamma$ lines, one half of the crystal is displaced in incremental steps of 1/96 (Secs.~\ref{sec:EAM}--\ref{sec:Tersoff}) and 1/12 (Secs.~\ref{sec:BOP} and \ref{sec:DFT}) of the corresponding periodicity distances, i.e., $a_0$\hkl[1-10] or $a_0/2$\hkl[1-11].
After each displacement step, the energy of the corresponding configuration was minimized using the relaxation algorithms and convergence criteria as described in the following (Secs.~\ref{sec:EAM}--\ref{sec:DFT}).
For both setup geometries, the GSFE $\gamma$ is calculated as the difference between the total potential energy of the shifted crystal $E^\text{tot}$ and the energy $E_0^\text{tot}$ of the initial, undisplaced, configuration \cite{Vit68}:
\begin{eqnarray}
\gamma & = & \frac{\left(E^\text{tot}-E_0^\text{tot}\right)}{A_\text{110}}
\end{eqnarray}
where $A_\text{110}$ is the area of the \hkl(110) plane in the simulation box.
In most cases, the $\gamma$ lines in the \hkl[1-11] Burgers vector direction were determined for comparison.

Atomistic configurations are always visualized using OVITO \cite{ovito,Stu10} and analyzed using the CNA \cite{Hon87} and the recently developed BDA \cite{Moe2016,bda}.

\subsection{Embedded Atom Method Potentials}
\label{sec:EAM}

According to the EAM formalism, the total potential energy $E^\text{tot}_\text{pot}$ of an $N$-atom configuration is given by \cite{Daw84}:
\begin{eqnarray}
 E^\text{tot}_\text{pot} & = & \sum \limits_{{i=0}}^{{N}} E_\text{pair}^\text{i} + \sum \limits_{{i=0}}^{N} E_\text{embed}^\text{i},
 \label{eq:EAM} \\
 E_\text{pair}^\text{i}  & = & \sum \limits_{{j=0}}^{{N}} \frac{1}{2} V(r_\text{ij}) \text{ for } j \neq i, \label{eq:EAM_pair} \\
 E^\text{i}_\text{embed} & = & F(\rho^\text{i}) = F \left( \sum \limits_{j=0}^N f(r_\text{ij}) \right) \text{ for } j \neq i.
 \label{eq:EAM_embed}
\end{eqnarray}
The radial-symmetric functions, $V(r_\text{ij})$ and $f(r_\text{ij})$, depend only on $r_\text{ij}$, i.e., the distance between two atoms $i$ and $j$ and $\rho^\text{i}$ is the electron density at the site of atom $i$.
The non-pairwise embedding energy term, $F(\rho^\text{i})$, describes the energy contribution arising from the electron density.

In this study, we compare a variety of EAM-type potentials for Fe spanning almost 30 years of potential development:
from the historic Finnis-Sinclair (FS) potential in the modified form of Marchese \etal{} \cite{Fin84,Mar88}, abbreviated as "MFS", to the most recent "Marinica11" potential by Proville \etal{} \cite{Rod12}.
In addition, we studied the FS potentials for V, Nb, Ta, Mo, and W in the modified version of Ackland and Thetford \cite{Ack87}, abbreviated as ATFS". 
Since such an Ackland-Thetford correction does not exist for Cr, we used the original FS potential instead. 
For Nb, Mo and W, the additional potential parameterizations by Fellinger \etal{} \cite{Fellinger2010}, Ackland \etal{} \cite{Han2003,Ackland2009}, Smirnova \etal{} \cite{Smirnova2013}, and Wang \etal{} \cite{Wang2014} were used as well.
All EAM potentials used in our study are listed in \Tref{tab:overview}.

In case of EAM potentials, energy minimization is performed using the software package IMD \cite{Bit01,imd} with the FIRE relaxator \cite{Bit06b} until the force-norm $\|F\|$ is below 10$^{-8}$ eV/\r{A}.

\subsection{Modified Embedded Atom Method Potentials}
\label{sec:MEAM}

As EAM potentials are unable to account for directional bonds, the MEAM has been developed by Baskes \etal \cite{Bas92} to account for this effect, which is specifically important for bcc metals.
As in the EAM formalism, the energy is still the sum of a distance-dependent pair-potential term and an embedding-energy term that depends on the electron density $\rho$.
Whereas in the EAM, the assumption was made that $\rho$ is a linear superposition of spherically averaged electron densities from all neighbor atoms, in the MEAMs $\rho$ has a more sophisticated form  that incorporates directional dependence in the electron density \cite{Tad11}.
The MEAM was extended to include second-nearest neighbor (2nn) interactions by Lee \etal{} \cite{Lee01}, which was shown to be an important feature to simulate the behavior of bcc metals more accurately.

We use the original 2nn-MEAM potential by Lee \etal{} \cite{Lee01}, which is also part of the Fe-Ti-C potential of Kim \etal{} \cite{Kim2009}.
In addition, two more recent parameter sets "MEAM-p" and "MEAM-T" by Lee \etal{} \cite{Lee12} were used. 
Both parameter sets have been used in the Fe parts of MEAM potentials for Fe alloys, namely for AlSiMgCuFe by Jelinek \etal{} \cite{Jel12} and for Fe-C by Liyanage \etal{} \cite{Liyanage2014}.

In case of MEAM potentials, energy minimization is performed using the software package LAMMPS \cite{Plimpton1995,lammps} with the FIRE relaxator \cite{Bit06b} until the force-norm $\|F\|$ is below 10$^{-6}$ eV/\r{A}. 

\subsection{Angular-Dependent Potentials}
\label{sec:ADP}

ADPs are an alternative approach to include directional bonding in the EAM formalism as proposed by Mishin \etal{} \cite{Mis05}.
In addition to the classical EAM functions [see Eqs.~(\ref{eq:EAM}--\ref{eq:EAM_embed}], they contain measures for the dipole and quadrupole distortions of the local atomic environment. 
The main difference from the MEAM formalism introduced above is that the higher-order multipoles contribute to the electron density in MEAM potentials whereas in the ADP method they contribute directly to the total energy.

Currently, ADPs are less common than MEAM potentials and available only for a few material systems \cite{Mis05,Mishin2006,Has08}. 
In this paper, we use the original ADP parameterization of Mishin \etal \cite{Mis05} for Fe. 
However, it should be noted here that for this parameterization the values for the elastic constants are only one half of the experimental ones \cite{MishinPC}.
It is therefore not a useful description whenever the elastic material response is of interest, as, e.g., for fracture.
For the ADP used here, energy minimization is performed using the same simulation details as for the EAM potentials described in Sec.~\ref{sec:EAM}. 

\subsection{Tersoff Potentials}
\label{sec:Tersoff}

An empirical approach based on the chemical concept of bond order was developed by Tersoff \cite{Tersoff1986}.
Tersoff potentials are based on Morse-type pair interactions \cite{Mor29} whose attractive parts include a function that depends on the number, strength, and angles of the interatomic bonds.

For Fe, two different Tersoff potentials were proposed, one by M\"uller \etal{} \cite{Mue07} and another by Bj\"orkas \etal{} \cite{Bjorkas2007}.
Note that the latter potential contains a slight modification of the repulsive part of the former to yield a more realistic description of the short-range behavior between atomic nuclei. 
This modification by Bj\"orkas \etal{} was also used to describe the Fe-Fe bonding in the Fe-Cr-C potential by Henriksson \etal{} \cite{Henriksson2013}. 
For both Tersoff potentials employed here, energy minimization is performed with the same simulation details as for the MEAM potentials described in Sec.~\ref{sec:MEAM}. 

\subsection{Bond-order Potentials}
\label{sec:BOP}

A common characteristic of all previously mentioned classical interatomic potentials is the lack of magnetic interactions. 
These potentials are therefore neither able to describe the broad variety of magnetic phases of Fe nor provide any information about local magnetic phenomena in the vicinity of crystal defects \cite{Mrovec2011}.
One of the most successful methods to overcome this shortcoming are bond-order potentials (BOPs) \cite{Pet89}.
This family of potentials is based on the tight-binding (TB) approximation and includes a description of both the electronic structure and magnetic interactions according to the Stoner theory of itinerant magnetism \cite{Sto39}.
For a recent review on the derivation and parameterisation of BOPs we refer the reader to Ref. \cite{Drautz2015}

BOPs represent a bridge between the classical potentials mentioned above and the more accurate but computationally more expensive DFT calculations described in the next subsection.  
For this reason, both small and large setup geometries, cf. \fref{fig:setups}, are compared using the "numerical" BOP for Fe by Mrovec \etal{} \cite{Mrovec2011} with the computational details (OXON code \cite{Horsfield1996}, relaxation with FIRE \cite{Bit06b} until the maximum force was less than 0.01 eV/\r{A}) as described in Ref.~\cite{Mrovec2011}.
The procedure is repeated for the small setup geometry using an analytical' BOP that Ford \etal{} \cite{Ford2014} devised from a TB model of Madsen \etal \cite{Madsen2011} within the framework of analytic BOPs for d-valent systems \cite{Drautz2006}.
In this case, the simulations are performed using the BOPfox code \cite{bopfox} with damped-Newtonian relaxation until the energy difference is below 0.001 eV/atom or the maximum force is below 0.001 eV/\r{A}.

\subsection{Density-Functional Theory Calculations}
\label{sec:DFT}
Most DFT calculations \cite{Hohenberg1964,Kohn1965} were carried out with the plane-wave code \textit{PWscf} of the {\sc Quantum Espresso} software package \cite{qe2009,quantumespresso} using the Perdew-Burke-Ernzerhof (PBE) exchange-correlation functional \cite{Perdew1996}, which is based on the generalized gradient approximation (GGA), and Vanderbilt-type ultrasoft pseudopotentials (USPP) \cite{Vanderbilt1990}.
DFT calculations with USPPs are expected to yield comparable results for GSFEs as compared with results for the more recent projector-augmented wave (PAW) methods; 
see, e.g., Ref. \cite{Ven10}.
The kinetic energy cutoff of the USPPs was 30 Ry for wave functions and 120 Ry for charge densities and potentials \cite{MeyerPC}.
To determine the properties of magnetic Fe, spin-polarized calculations were performed.  
The convergence threshold for self-consistent calculations was 10$^{-8}$ Ry (approximately 1.36$\times 10^{-7}$ eV).
Shifted 39$\times$3$\times$39 \textit{k}-point meshes for Brillouin-zone integrations were generated by the Monkhorst-Pack scheme \cite{Monkhorst1976} and the fractional occupations of the electronic states were realized by a Gaussian broadening \cite{Fu1983}. 
Atomic positions were relaxed by minimizing the atomic forces using the Broyden-Fletcher-Goldfarb-Shanno (BFGS) relaxation scheme \cite{Press2007} with a convergence threshold for the largest residual force component of 10$^{-3}$ Ry/Bohr (approximately 2.6$\times 10^{-2}$ eV/\r{A}).

To check the independence of the DFT results from the used simulation package and realizations of the external potential,
we also performed DFT calculations using the Vienna Ab initio Simulation Package (VASP) \cite{Kresse1993,vasp}.
The simulation details used with VASP (GGA, PBE, and PAW \cite{Kresse1999}) are reported in more detail in Ref.~\cite{Bleskov2016}. 
The determined $\gamma$ lines were in good agreement with other calculations using the VASP code and reference data from literature \cite{Ven10}.

\begin{table}
\caption{\label{tab:overview} 
Overview of material models for bcc metals and presence (\yes) or absence (\no) of a local \hkl(110) GSFE minimum 
min($\gamma_\text{[hkl]}$) where [hkl] is the shearing direction; 
US, uniaxial strain in \hkl[110] direction;
EBS, equi-biaxial strain in \hkl[110] and \hkl[1-10] direction; 
$\rightarrow$, largest strain studied. }
\scriptsize
\begin{tabular}{@{}cc|c|c|c|c|c}
\hline\hline
Element & Ref. & min($\gamma_{[1\bar{1}1]}$) & \multicolumn{4}{c}{min($\gamma_{[1\bar{1}0]}$)} \\
(Potential)&  & $\varepsilon$=$0\%$ & $\varepsilon$=$0\%$ & $\varepsilon_\text{US}$=$7.5\%$ & $\varepsilon_\text{EBS}$=$5\%$ & $\varepsilon_\text{EBS}$=$7.5\%$  \\
\hline
\multicolumn{7}{c}{\textit{Embedded Atom Method (EAM) Potentials}}\\
\hline
Fe (Mendelev-II)  & \cite{Men03} & 					\yes & 	\no & 	\yes &	\yes & \yes \\
Fe (Chamati)  & 	\cite{Cham06} & 					\no & 	\no & 	\no &	 \no & \yes \\
Fe (Chiesa) & 		\ocite{Dud05,Chi11} & 		\yes &	\no &	\no & \yes & \yes\\
Fe (MFS)  & 		\ocite{Fin84,Mar88} & 			\no &	\no & 	\no & 	\no & \yes \\
Fe (Simonelli)  & 	\cite{Sim93} & 				\no &	\no & 	\no & 	\no & \yes\\
Fe (Men-IIext) &	\cite{Ack04} & 					\no & 	\no & 	\yes & 	\yes & \yes \\
Fe (Gordon)  & 		\cite{Gor11} & 				\no &	\no & 	\no & 	\no & \yes\\
Fe (Marinica07)  & 	\cite{Mal10} & 				\no &	\no & 	\no & 	\yesno \footnote[1]{original bcc lattice transformed to hcp lattice} & \yesno \footnotemark[1] \\
Fe (Marinica11)  & 	\cite{Rod12} & 				\yes &	\no & 	\yes & 	\no & \yes	\\
V (ATFS)  & 		\ocite{Fin84,Ack87} & 			\yes &	\no & 	{---} &	\no & \yes  \\
Nb (ATFS)  & 		\ocite{Fin84,Ack87} & 		\yes &  \no & 	{---} & \no & \yes \\
Nb (Fellinger)  & 	\cite{Fellinger2010} & 		\no & 	\no & {---} & \yes & \yes\\
Ta (ATFS)  & 		\ocite{Fin84,Ack87} & 		\no & 	\no & 	{---} & \no & \yes \\
Cr (FS)  & 			\cite{Fin84} & 				\no & 	\no & 	{---} &	\no & \yes \\
Mo (ATFS)  & 		\ocite{Fin84,Ack87} & 		\no & 	\no & 	{---} & \yes & \yes\\
Mo (Ackland)  & 	\ocite{Han2003,Ackland2009}&		\yes & 	\yes &	{---} & \yes & \yes\\
Mo (Smirnova)  & 	\cite{Smirnova2013} & 			\no & 	\no & 	{---} & \yes & \yes\\
W (ATFS)  & 		\ocite{Fin84,Ack87} & 			\no & 	\no & 	{---} & \no & \yes\\
W (Ackland) & 		\ocite{Han2003,Ackland2009}&	\yes & 	\yes & 	{---} & \yes & \yes\\
W (Wang) & 			\cite{Wang2014} & 				\no &	\no & {---} & \no & \yes\\
\hline
\multicolumn{7}{c}{\textit{Modified Embedded Atom Method (MEAM) Potentials}}\\
\hline
Fe (Lee2001) & 		\cite{Lee01} & 				\no & 	\no & 	{---} &	\no & ---\\
Fe (Lee2012-p) &	\cite{Lee12} & 					\no & 	\yes & 	{---} &	\yesno \footnotemark[1] & ---\\
Fe (Lee2012-T) &	\cite{Lee12} & 					\no & 	\no & 	{---} &	\no & ---\\
\hline
\multicolumn{7}{c}{\textit{Angular-Dependent Potential (ADP)}}\\
\hline
Fe (Mishin) & 		\cite{Mis05} & 				\no 	& \no & {---} &	\no & ---\\
\hline
\multicolumn{7}{c}{\textit{Tersoff Potentials}}\\
\hline
Fe (M\"uller) & 		\cite{Mue07} & 				\no & \no & {---} & \yes & ---\\
Fe (Bj\"orkas) & 	\cite{Bjorkas2007} & 			\no & \no & {---} & \yes & ---\\
\hline
\multicolumn{7}{c}{\textit{Bond-Order Potentials (BOP)}}\\
\hline
Fe (Mrovec) & 		\cite{Mrovec2011}		& 		\no &	\no & {---} & \no & ---\\
Fe (Ford) & 		\ocite{Madsen2011,Ford2014}	& 	\no & 	\no & {---} & \no & ---\\
\hline
\multicolumn{7}{c}{\textit{Density-functional theory (DFT) calculations}}\\
\hline
Fe (PAW-PBE) &		\ocite{Kresse1993,vasp} & 	\no & 	\no &	\no~$(\rightarrow 10\%)$ &	\no & ---\\
Fe (USPP-PBE) &	 	\cite{MeyerPC} & 				\no & 	\no &	\no~$(\rightarrow 10\%)$ &	\no & \no$(\rightarrow 22.5\%)$\\
Nb (USPP-PBE) &		\cite{MeyerPC} & 				{---}& 	\no &	{---} &	\no & ---\\
Ta (USPP-PBE) & 	\cite{quantumespresso} & 		{---}& 	\no &	{---} &	\no & ---\\
Mo (USPP-PBE) & 	\cite{MeyerPC} & 				{---}& 	\no &	{---} &	\no & ---\\
W (USPP-PBE) & 		\cite{MeyerPC} & 				\no & 	\no &	{---} &	\no & ---\\
\hline\hline
\end{tabular}
\end{table}

\section{Results}
\label{sec:results_discussion}

\begin{figure*}
\includegraphics[width=\textwidth]{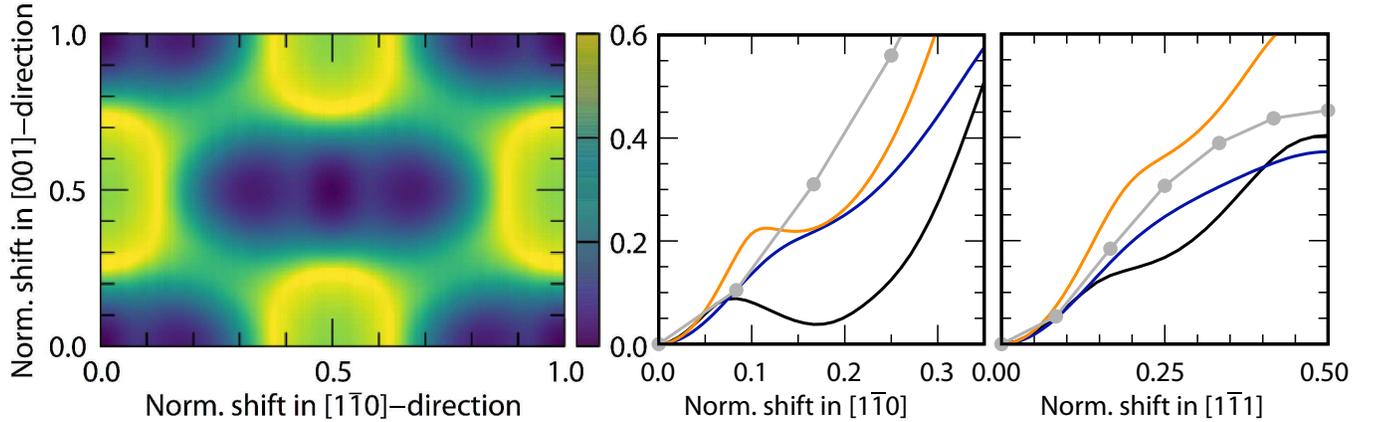}
\caption{$\gamma$ surfaces of the \hkl(110) glide plane and corresponding 1D GSFE line profiles ($\gamma$ lines) in \hkl[1-10] and \hkl[1-11] directions  as function of external loads:
(a) no external load;
(b) 7.5~\% uniaxial strain (US);
(c) 5~\% equi-biaxial strain (EBS).
Note that the scale of the $\gamma$ lines is the same as for the colorbars of the $\gamma$ surfaces.   
The shifts in \hkl[001], \hkl[1-10], and \hkl[1-11] directions are normalized, i.e., divided by the corresponding periodicity distances $a_0$, $\sqrt{\text{2}}a_0$, and $\sqrt{\text{3}}/2a_0$.
The $\gamma$ surfaces are displayed for the Mendelev-II potential \cite{Men03}.
The $\gamma$ lines are plotted for the Mendelev-II \cite{Men03}, Chamati \cite{Cham06}, and Chiesa \cite{Chi11} potentials as well as for DFT calculations.
All GSFE curves are plotted with respect to the energy of the crystals at zero normalized shifts irrespective of their strain state.
The Mendelev-II and Chiesa potentials tend to result in a local GSFE minimum along the \hkl[1-10] direction under external strain [(b) and (c)];
this behavior is observed neither for the Chamati potential nor in DFT calculations. 
}
\label{fig:gsfe_Fe} 
\end{figure*}

\begin{figure*}
\includegraphics[width=\textwidth]{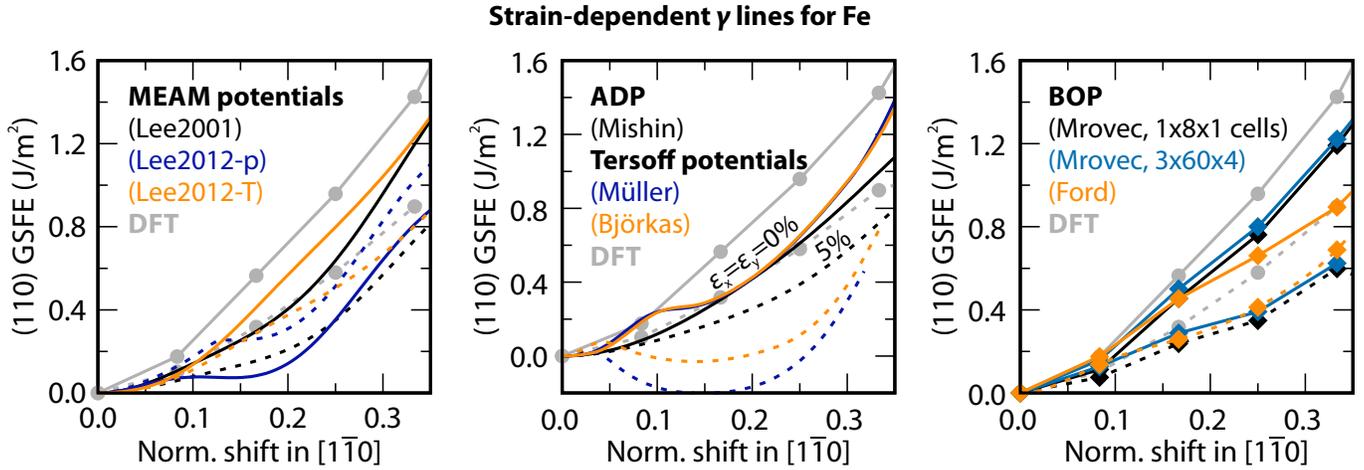}
\caption{Strain-dependent $\gamma$ lines for Fe using MEAM, ADPs, and BOPs in comparison to the DFT (USPP) data.
The GSFE data are plotted for 0\% (solid lines) and 5\% (dashed) applied equi-biaxial strain parallel to both the \hkl(110) plane normal and the \hkl[1-10] shearing direction. 
The formation of a local GSFE minimum or at least a terrace at normalized shifts around 0.15 as observed for several EAM potentials (Figs. \ref{fig:gsfe_Fe} and \ref{fig:gsfe_bcc}) is only observed for the Lee2012-p MEAM potential.
In this case, the GSFE exhibits generally higher values for 5~\% than for 0~\% applied strains, which is attributed to a bcc $\rightarrow$ hcp phase transformation for the unshifted configuration at 5~\% applied strain.
}
\label{fig:gsfe_Fe_MEAM-ADP-BOP} 
\end{figure*}

\begin{figure*}
\includegraphics[width=\textwidth]{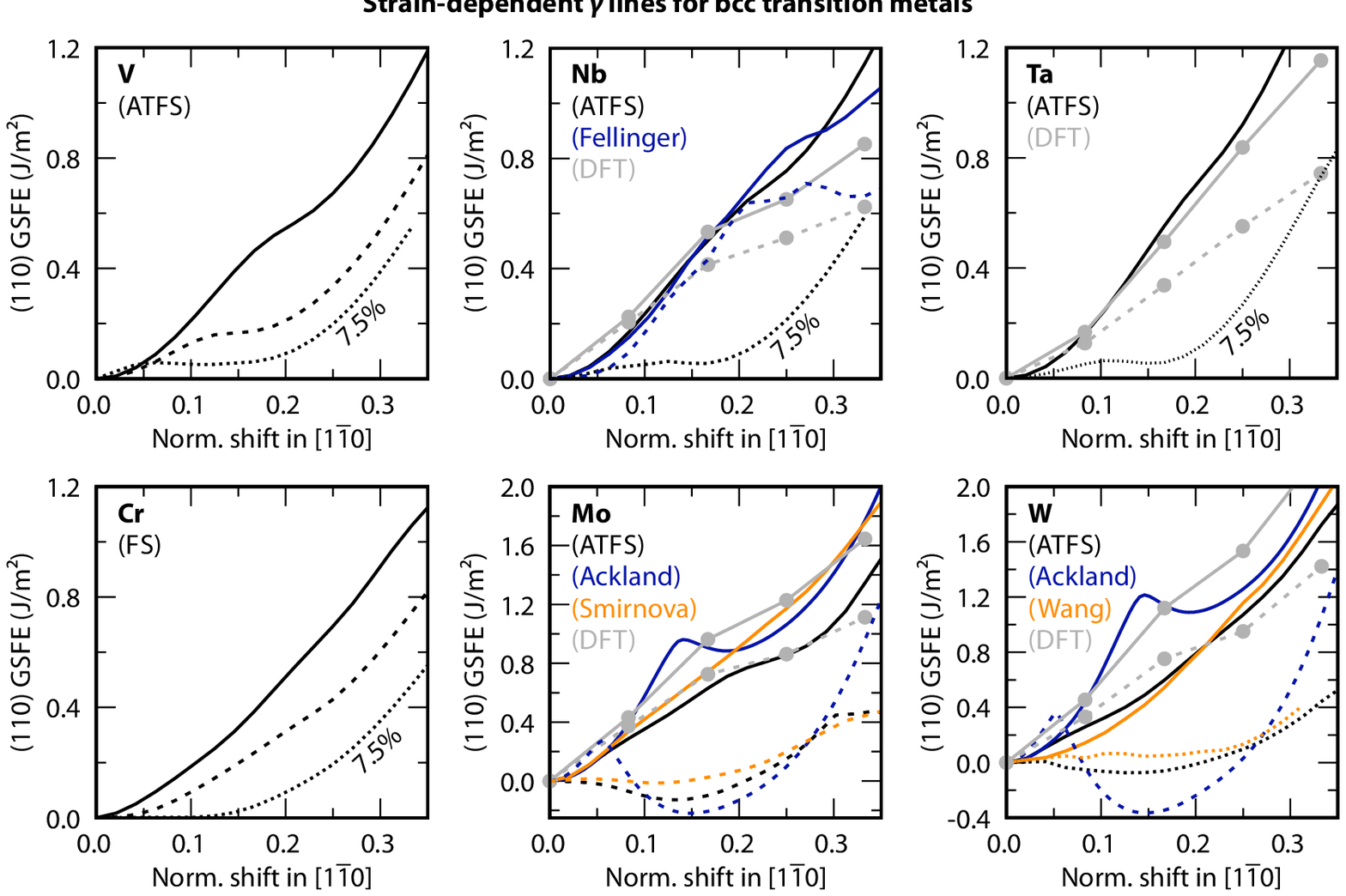}
\caption{Strain-dependent $\gamma$ lines of various EAM potentials for the bcc metals V, Nb, Ta, Cr, Mo, and W.
The GSFE data are plotted for 0\% (solid lines), 5\% (dashed), and 7.5\% (dotted) applied equi-biaxial strain parallel to both the \hkl(110) plane normal and the \hkl[1-10] shearing direction. 
The shift in \hkl[1-10] direction is normalized, i.e., divided by the corresponding periodicity distance $\sqrt{\text{2}}a_0$.
Relaxed stacking fault energies for Nb, Ta, Mo, and W were additionally calculated using DFT.
}
\label{fig:gsfe_bcc} 
\end{figure*}

The results for the occurrence of strain-dependent local minima in the calculated $\gamma$ lines are summarized in \tref{tab:overview}.
For reference, we always compared the unstable stacking fault energies $\gamma_\text{usf}$ along the \hkl[1-11] Burgers vector direction under unstrained conditions with other published data for the corresponding interaction models.
As it is indicated in the table, some of the EAM potentials exhibit a local minimum in the middle of the $\gamma$ line in \hkl[1-11] direction [denoted as min($\gamma_{[1\bar{1}1]}$)].
This leads to a "double hump" shape of the GSFE where there should be only one maximum (according to the predictions of the more sophisticated methods).
The Gordon potential \cite{Gor11}, for instance, was developed exactly with the purpose to suppress this unwanted feature of many EAM potentials. 
The remaining three columns of \tref{tab:overview} indicate whether there is a local minimum in the corresponding $\gamma$ line under zero strain (ZS), 7.5\% uni-axial strain (US), and/or 5\% equi-biaxial strain (EBS)  or not.
For Fe, the size independence of the results was ensured by using simulation box sizes containing 480, 2,880, and 5,760 atoms for the Mendelev-II EAM potential, 16 and 2,880 atoms for the Mrovec BOP, and 16 and 32 atoms in case of DFT calculations.

In \fref{fig:gsfe_Fe} the $\gamma$ surfaces of Fe are shown under unstrained (a) as well as for 7.5\% uniaxial strain in \hkl[110] (b) and 5\% biaxial strain in the \hkl[1-10] and \hkl[110] directions (c). 
Note that all GSFE data points are plotted with respect to the energy of the crystals at zero normalized shifts irrespective of their strain state.
This conceals the fact that all configurations under strained conditions are of course of higher energy than those under unstrained conditions, 
but it allows one to compare  directly their relative evolution on the same scale as also displayed in Figs.~\ref{fig:gsfe_Fe_MEAM-ADP-BOP} and \ref{fig:gsfe_bcc}. 
While the $\gamma$ surfaces are only shown for the popular Mendelev-II potential \cite{Men03}, the $\gamma$ lines are also presented for the more recent Chamati \cite{Cham06} and Chiesa potentials \cite{Chi11}.
As expected from literature \cite{Ven10}, the DFT results obtained with USPPs are in good agreement with the results for the more recent PAW method (see Supplemental Material).

It can clearly be seen that both the Mendelev-II and Chiesa potentials lead to local minima at normalized shifts in the range of 0.15 and 0.20 in the \hkl[1-10]-direction.
For the strain states shown in the figure, this was not the case for the Chamati potential and the DFT calculations.
It is important to note that the shape of the GSFE curves for the Mendelev-II and Chiesa potentials differs qualitatively from the shape of the Chamati potential already at 0\% strain [see \fref{fig:gsfe_Fe}(a)].
The GSFE curves of the Mendelev-II and Chiesa potentials exhibit a consecutive change of curvature from positive to negative and back,
which leads to a shoulder-like shape of the GSFE curve.
By comparing the GSFE curves in Figs. \ref{fig:gsfe_Fe}(b) and \ref{fig:gsfe_Fe}(c), it can be seen that this "shoulder" often develops into a "hump" for applied strains.
This "hump," which is identical to the formation of a local minimum, is shifted to smaller values of normalized shifts with increasing strain.
Interestingly, this shoulder-like shape is also observed for relative shifts in the usual Burgers vector direction (\hkl[1-11] direction, rightmost subfigures in \fref{fig:gsfe_Fe}).
The Chamati potential, on the other hand, does initially not show such a shoulder, but starts to develop a similar shape at applied strains; see Figs. \ref{fig:gsfe_Fe}(b) and \ref{fig:gsfe_Fe}(c).

As it can be seen in \tref{tab:overview} and in \fref{fig:gsfe_Fe}, the GSFE curves calculated with DFT methods do not develop a local minimum even for EBS values up to 10\% whereas a local GSFE minimum occurs for the Chamati potential when the EBS was increased to 7.5\%. 
The Chamati potential shares this behavior with all other EAM potentials. 
From the more sophisticated material models only the MEAM potential by Lee \etal{} ("Lee2012-p") and the Tersoff potentials by M\"uller \etal{} and Bj\"orkas \etal{} showed local minima (see \fref{fig:gsfe_Fe_MEAM-ADP-BOP}).
In the latter two cases, the minimum is not only local but \textit{global}, which indicates the instability of the bcc phase under these conditions. 
For the other MEAM potentials as well as the ADP and BOPs local GSFE minima were not observed in the studied range of applied strains.

Figure \ref{fig:gsfe_bcc} summarizes the resulting $\gamma$ lines in the \hkl[1-10] direction of the EAM potentials for the other bcc transition metals V, Nb, Ta, Cr, Mo, and W.
In case of Nb, Ta, Mo, and W the results for the EAM potentials were compared to DFT calculations. 
Note that the EAM potentials for Cr (FS) and Mo (ATFS)
show a completely different behavior for $\varepsilon_\text{\hkl[1-10]}=\varepsilon_\text{\hkl[110]}=5\%$.
While the Mo potential shows a local (and even global minimum), no such minimum is observed for Cr.
The non-physical predictions of the EAM potentials for Mo are in this case again confirmed by DFT calculations for the same strain state.
 
\section{Discussion}
\label{sec:discussion} 

Our results indicate that the observation of \hkl{110} planar faults---sometimes 
discussed as "fcc formation", see e.g. Refs. \cite{Lat03,Bor11,Vat11,Vatne2011,Ers12,Skogsrud2015,Zhang2012a}---is linked to the appearance 
of a local GSFE minimum under an applied strain.
Zhang \etal{} already emphasized that the correct shape of GSFE curves is a challenge to empirical potentials \cite{Zhang2011e}.
In their work, the authors compared the GSFE curves for the \hkl{112} twin plane in the classical \hkl<111> Burgers vector direction for five different 
interatomic potentials and DFT calculations for Nb. 
Similar to our work, some of the potentials lead to local GSFE minima at $x \approx 1/3$ and 2/3 
(where $x$ corresponds to the shift in \hkl<111> direction normalized to the Burgers vector length $a_0/2\hkl<111>$) for applied external strains. 
They showed that the trend to give a local GSFE minimum at applied strains correlates with the trend to give a local GSFE minimum 
in the \hkl<111> Burgers vector direction but at $x = 1/2$ already at zero strain, leading to a double-hump shape of the GSFE.
This observation is in contrast to our results where the focus is on the GSFE in \hkl<110> directions (and on \hkl{110} planes) and where 
such a straightforward correlation is absent, cf. \tref{tab:overview}.
In other words, the formation of \hkl{110} PFs under applied strains is found to be independent of a double-hump GSFE at zero strain. 
Moreover, we applied comparably moderate \textit{tensile} stresses around 20 GPa (for Fe)
instead of 50 GPa \textit{compressive} strains in the work by Zhang \etal{}
In our work, the ATFS \cite{Fin84,Ack87} and Fellinger \cite{Fellinger2010} potentials, 
which Zhang \etal{} found to reproduce the GSFE at applied strains for the \hkl{112} plane in the classical \hkl<111> shearing direction, 
failed to do so for the \hkl{110} plane in \hkl<110> shearing directions which are typically not subject to intense research efforts.
It is therefore important to note, that an analysis of possible reasons for the occurrence of local GSFE minima at applied strains is currently lacking.

\begin{figure}
\includegraphics[width=\columnwidth]{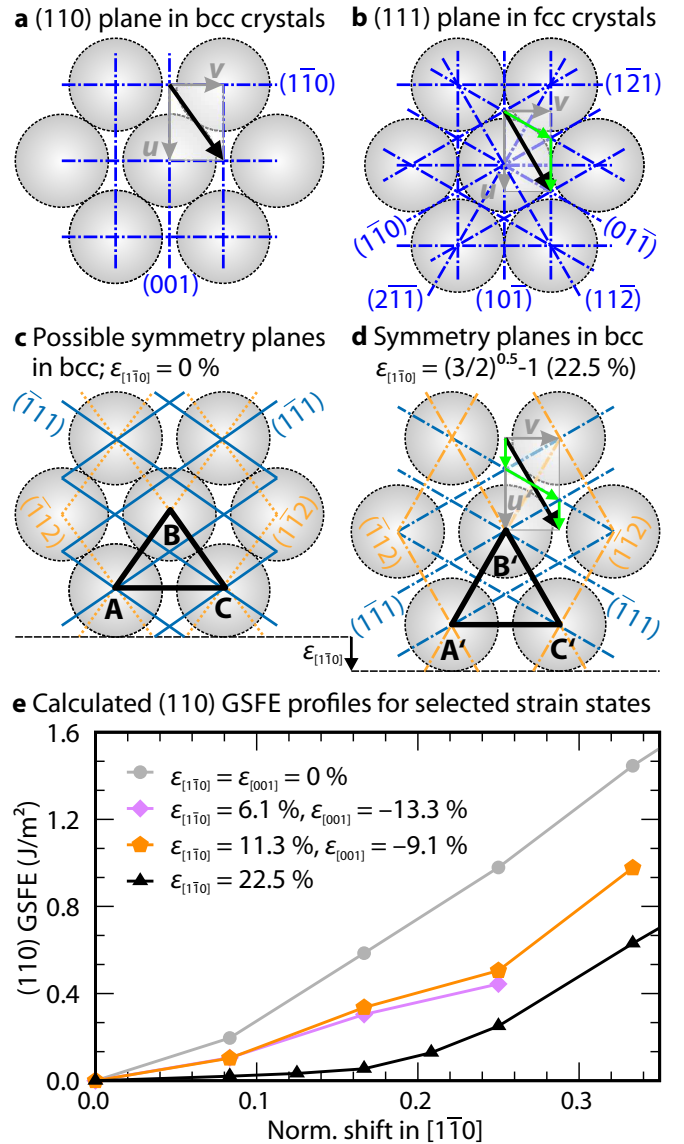}
\caption{Symmetry planes and Burgers vectors (black solid arrows) on the \hkl(110) slip plane in bcc (a) and the \hkl(111) slip plane in fcc crystals (b).
Under zero-strain conditions, the \hkl{111} and \hkl{112} planes are no mirror planes of bcc crystals (c).
Only if the crystal is strained in such a way that the isosceles triangle ABC becomes the equilateral triangle A'B'C' (d),
then the \hkl{111} and \hkl{112} planes are mirror planes, form additional intersections, and the appearance of a local GSFE minimum is theoretically possible.
The 1D GSFE profiles as calculated with DFT (e) for this (black symbols) and other extreme scenarios associated with an equilateral triangle A'B'C', however, do not predict a local GSFE minimum.
}
\label{fig:symmetry}
\end{figure}

From a theoretical point of view, the occurrence of a local GSFE minimum is not expected in bcc materials.
According to Neumann's principle \cite{Neumann1885}, which states that the symmetry of physical properties is linked to the symmetry of the crystal, 
a local GSFE minimum (i.e. zero slope) would necessitate the crossing of two non-parallel mirror planes perpendicular to the studied glide plane.
For bcc crystals at zero applied strains, this is clearly not the case while it is for fcc materials cf. Figs. \ref{fig:symmetry}(a) and \fref{fig:symmetry}(b).
However, under conditions of extreme strains (e.g. $\varepsilon_{\hkl[1-10]} = 22.5\%$) the \hkl{111} and \hkl{112} planes can become additional 
mirror planes of bcc crystals, [cf. Figs. \ref{fig:symmetry}(c) and \fref{fig:symmetry}(d)], thereby enabling the theoretical possibility of a local GSFE minimum.
As it can be seen in the figure, only strains in the \hkl[1-10] and \hkl[001] directions are needed for such a scenario, i.e., 
no strain perpendicular to the \hkl(100) PF plane is needed. 
On the other hand, some EAM potentials presented in Sec. \ref{sec:results_discussion} predict a GSFE minimum even if no strains are applied in these lateral directions but only in the \hkl[110] direction.
We therefore conclude that the appearance of a local GSFE minimum does not occur for reasons of symmetry but is an unrealistic artifact rooted in the 
oversimplifications assumed to construct some of the material models.
In particular, the strain-dependent formation of a local GSFE minimum was not observed for BOPs and in DFT calculations while it is predicted by all EAM and Tersoff potentials as well as for one MEAM potential (Lee2012-p).
We further note that even for $\varepsilon_{\hkl[1-10]} \approx 22.5\% $ and other scenarios of extreme applied strains, 
DFT calculations do not predict a local GSFE minimum as displayed in \fref{fig:symmetry}(e).

\begin{figure}
\includegraphics[width=\columnwidth]{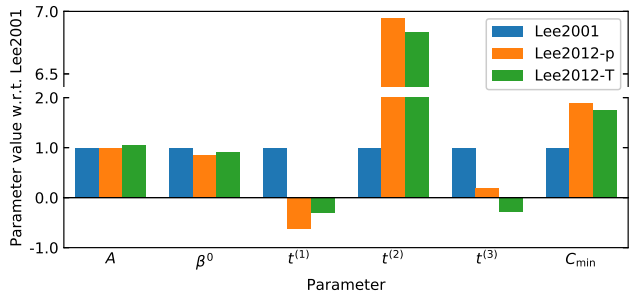}
\caption{Comparison of the parameters for the three MEAM potentials. 
The tendency to show a local GSFE minimum (in case of the Lee2012-p potential) correlates with the $C_\text{min}$ and $t^\text{(2)}$ parameters
and anti-correlates with the  $A$ and $\beta^0$ parameters.}
\label{fig:meam_parameters}
\end{figure} 

\begin{figure*}
\includegraphics[width=\textwidth]{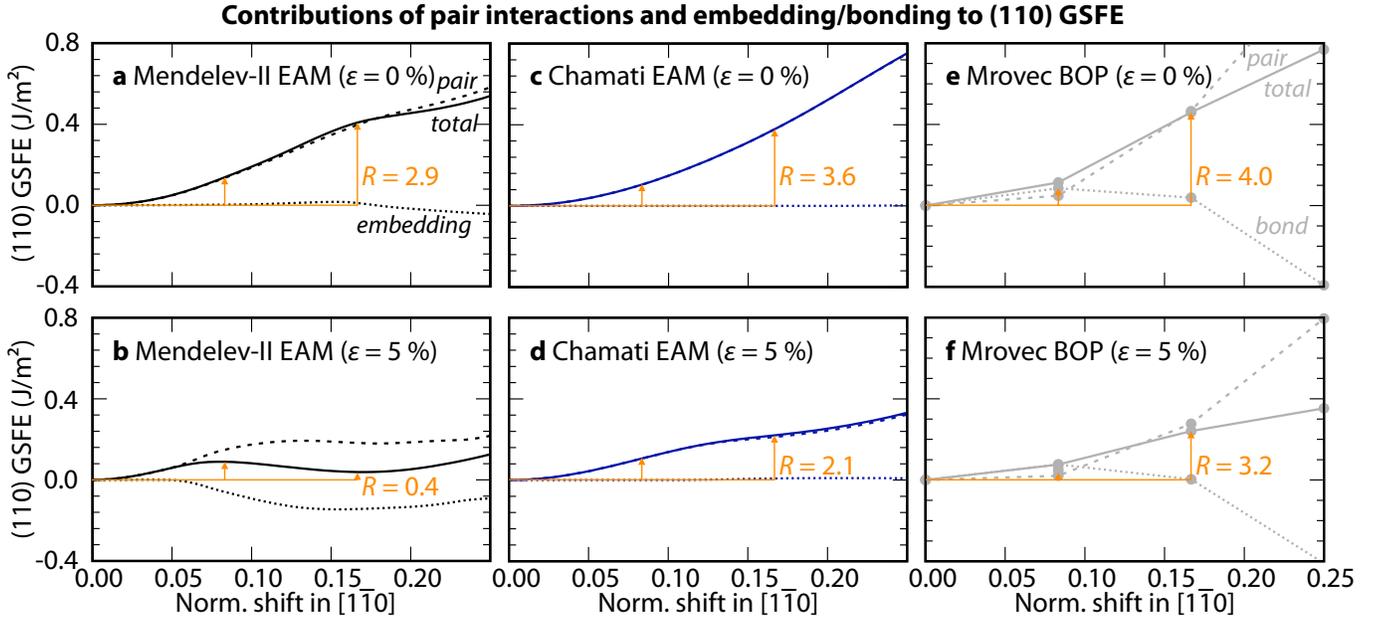}
\caption{\hkl(110) GSFE lines in \hkl[1-10] direction decomposed in contributions of pair/embedding energy (EAM potentials) and
pair/bond energy (BOP): (a) and (b), Mendelev-II EAM potential; (c) and (d), Chamati EAM potential; (e) and (f), Mrovec BOP. 
The top (bottom) row shows the decomposed $\gamma$ lines at 0\% (5\% equi-biaxial) strain.
See \eref{eq:R} for the definition of $R$. }
\label{fig:decompose_gsfe}
\end{figure*}

\begin{figure}
\includegraphics[width=\columnwidth]{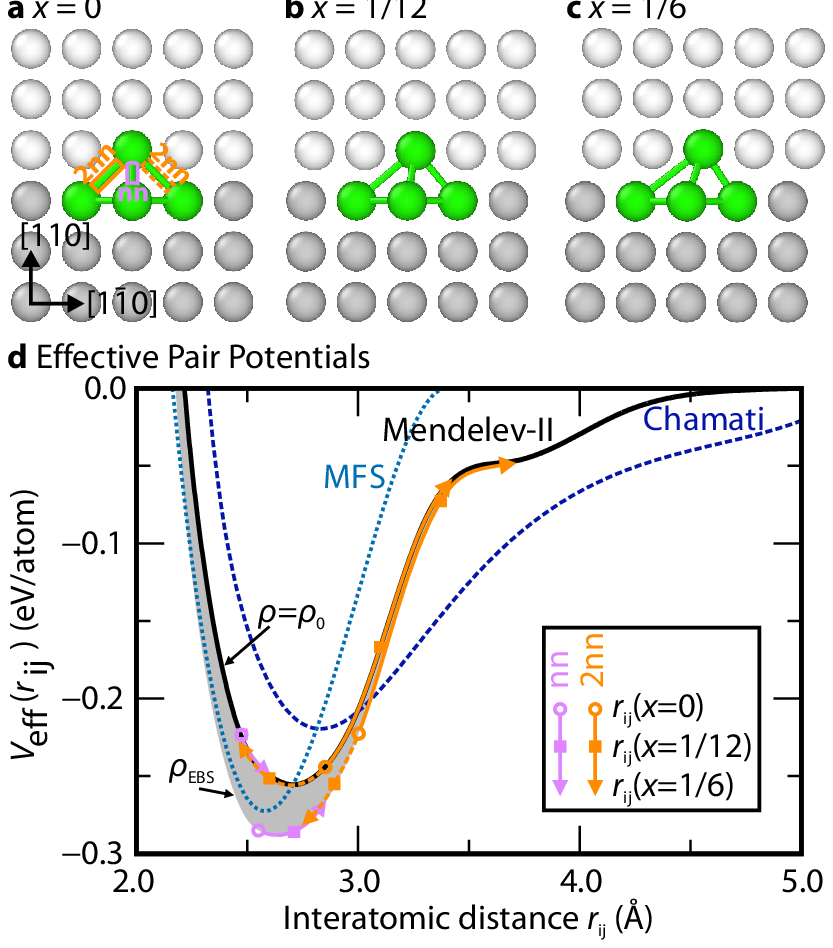}
\caption{Close-up views of the atomistic configurations at $x = 0$ (a), $x = 1/12$ (b), and $x = 1/6$ (c) for the Mendelev-II potential for Fe.
For one atom in the upper half crystal, nn and 2nn neighbor atoms and the bonds between them are indicated in green color.
The energy contributions due to the different neighbor distances are qualitatively compared using the effective pair potential ($V_\text{eff}$) format (d).
For comparison, $V_\text{eff}$($r_\text{ij}$) is also shown for the MFS and Chamati potentials, which are less prone to show a local GSFE minimum than the Mendelev-II potential (see \tref{tab:overview}).
For the Mendelev-II potential, the gray-shaded area indicates the change of $V_\text{eff}$ for decreasing electron densities (corresponding to
increasing applied strains).
The nn and 2nn bonds and their evolution upon increasing shifts $x$ are shown with purple and orange lines, respectively, for zero strain (electron density $\rho_0$)
and 5\% EBS ($\rho_\text{EBS}$).
While one 2nn bond is compressed [marked with a dashed line; see (a)], the other is stretched (solid line).
}
\label{fig:closeup}
\end{figure}

Why do certain models predict a local GSFE minimum and others do not?
Although the statistical basis of our results is clearly not large enough to give a conclusive answer to this question, 
we can shed some light on it by analyzing the contributions that are explicitly included in the different material models (see \tref{tab:summary_models}).
Together with the results presented in \tref{tab:overview}, this overview suggests that for obtaining an artifact-free GSFE surface the inclusion of direction-dependent
bonding is a necessary---but not sufficient---condition.
It is necessary since none of the EAM-type potentials (without direction dependence) results in an artifact-free behavior while some of the MEAM potentials (with direction dependence) do so. 
At the same time, it is only a sufficient condition since one MEAM and both Tersoff potentials (which both include direction dependencies) predict a local GSFE minimum.  
According to this analysis, it is not necessary to include electronic, magnetic, or bond-order effects to obtain a GSFE curve without a local minimum. 

The failure of purely radial-symmetric functions to correctly describe the GSFE can generally be understood by the insufficient description of the nearest neighbor bonds in the EAM formalism when the atomic environment is deformed and difficulties in modeling situations where atoms leave or come within the interaction range of EAM potentials as, e.g., at surfaces, along transformation paths such as the Bain path, and in the case of GSFE surfaces.
In a perfect crystal the contribution of further neighbor bonds is typically small (because the bond-order matrix elements of the further neighbors are significantly reduced). 
If, however, nearest neighbor atoms would be removed, the bond-order with further distant neighbors would increase. 
As the dependence of the strength of a bond on its local environment is not contained in the EAM, in order to suppress the 
formation of strong further distant bonds, EAM potentials have effectively to be cut off beyond the first or second neighbor shell of the equilibrium volume.
Furthermore, the EAM formalism neglects the fact that the bond order may vary significantly between nearest neighbour bonds if an atom is not coordinated homogeneously.

It is also important to note that the material models which formally include direction dependence but give a local GSFE minimum,
namely the MEAM and Tersoff potentials, still strongly rely on the contribution of pair potentials, 
e.g., of the Morse (Tersoff potential by M\"uller \etal \cite{Mue07}) or the Lennard-Jones type (ADP by Mishin \etal \cite{Mis05}).
For both model types at least 10 parameters define the characteristics of the potential and it is difficult to deduce which of these 
is most responsible for the formation of a local GSFE minimum.  
However, the result that only the Lee2012-p MEAM potential\cite{Lee12} gives a local GSFE minimum opens the possibility to compare the changes made 
to its parameters with those parameters of the Lee2001 \cite{Lee01} and Lee2012-T \cite{Lee12} potentials which do not give a local GSFE minimum.
Figure \ref{fig:meam_parameters} compares the parameters that differ between these three MEAM potentials.
From this comparison it can be seen that the tendency to lead to a local GSFE minimum correlates with the trend in the $C_\text{min}$ and $t^\text{(2)}$ parameters
and anti-correlates with the $A$ and $\beta^0$ parameters.
This agrees well with the statement of Lee and Baskes \cite{Lee01} that the effect of changes in $A$, $\beta^0$ and $C_\text{min}$ on the elastic constants and the energy differences between bcc, fcc, and hcp can be significant.
The analysis is, however, dramatically complicated by the distinct behavior of the Lee2012-p potential, which transforms from bcc $\rightarrow$ hcp already in the unshifted configuration at 5\% equi-biaxial strain and gives an \textit{increase} of the GSFE with applied strains, see \fref{fig:gsfe_Fe_MEAM-ADP-BOP}.
This transformation, however, is most likely not an artifact but due to its designation to reproduce the bcc-to-hcp transition pressure around 12 GPa (see Ref.~\cite{Lee12}).
For this reason, we can currently not propose a criterion for the tendency to form \hkl{110} PFs based on MEAM parameters.

We have also analyzed the training data that were used for parameterizing interatomic potentials as the training data generally determine the application range of a (material) model and it is therefore not \textit{a priori} clear how it will respond to "unknown" scenarios.
Due to the limited intrinsic transferability (as compared to BOP and DFT), the predictive power of classical potentials depends critically on the training set used during their parameterization.
We noticed a vague trend that the inclusion of elastic constants, surface energies, and phase stabilities at the same time (this is the case for the Chamati-EAM, Lee2001-MEAM, and Mishin-ADP as well as for the BOPs) may be sufficient but not necessary (the MFS potential, for instance, was only fitted to elastic constants) for reducing the tendency to form a local GSFE minimum; see Supplemental Material for details.

Coming back to our initial statement about the importance of direction dependence, we now ask whether distance-dependent potential functions are really not sufficient to predict the correct shape of the strain-dependent GSFE surface or
if the cutoff radii of the studied potentials (or their contributions due to pair interactions) were just too small.
A too small cutoff radius may disable important interactions between atoms across the stacking fault plane when they are shifted relative to each other or 
when they are pulled apart by the application of external strains.
However, the potentials with the smallest cutoff radius (MFS potential, $r_\text{cut} = 3.4$~\r{A}, including only first- and second-nearest neighbor interactions) 
and the largest cutoff radius (Chamati potential, $r_\text{cut} = 5.67$~\r{A}, including fifth-nearest neighbor interactions) both exhibit a similar 
behavior (local GSFE minimum at 7.5\% equi-biaxial strain; see \tref{tab:overview}).
Therefore---although we cannot ultimately rule out this possibility---our results indicate that a too small cutoff radius is not responsible for the formation of a local GSFE minimum.
A further indication for this conclusion is that third-nearest neighbor interactions are completely screened in the "Lee2012-T" MEAM potential \cite{Lee12} 
for which no local GSFE minimum is formed. 

An alternative explanation for the failure of EAM potentials to predict the correct shape of GSFE surfaces may be found in the decomposition 
of their total energy into contributions by "pair" and "embedding" energies.
This distinction, which dates back to the early days of the EAM formalism \cite{Daw84,Daw89}, 
is somewhat arbitrary in more recently developed potential where potentials are obtained by the parameterizing, e.g., cubic spline functions; 
see, for instance, the Fe potentials in Refs. \cite{Sim93,Men03,Gor11,Mal10,Rod12,Dud05,Chi11}.
In \fref{fig:decompose_gsfe} we compare the evolution of the "pair" and "embedding" contributions in dependence of the relative shift $x$ 
for two interatomic potentials (Mendelev-II and Chamati) and the BOP by Mrovec \etal{} for which the "bond" energy is plotted instead of the 
"embedding" energy. 
It can be seen that for all potentials the increase in total energy is mainly determined by pair interactions at zero strain and 
that the contribution of embedding/bond energy is negligible until approximately $x=1/6$.
When strains are applied (bottom row), this behavior persists only for the Chamati potential and the BOP [Figs. \ref{fig:decompose_gsfe}(d) and \ref{fig:decompose_gsfe}(f)]. 
For the Mendelev-II potential, on the other hand, the total energy is lowered due to the increased (negative) contribution of the embedding energy [\fref{fig:decompose_gsfe}(b)] leading to the formation a local minimum. 
This observation suggests that an increasing contribution of the embedding energy under applied strains could be responsible for the formation of \hkl{110} PFs.
As the individual contributions of pair/embedding energy depend, however, on the specific potential, this statement is not generalizable.

Under the assumption that first- and second-nearest neighbor interactions mainly determine the shape of the GSFE (as we have discussed before),
we will now focus on their interactions across the sheared \hkl(110) plane.
As we can see in Figs. \ref{fig:closeup}(a)--\ref{fig:closeup}(c), each atom is interacting with two first-nearest neighbors (nn) and two second-nearest neighbors (2nn). 
Upon shifting the two half crystals, both nn bonds and one of the 2nn bonds are stretched while the other 2nn bond is compressed. 
For analyzing the energy contributions of these interactions, we make use of the effective pair potential format \cite{Gor07,Joh89},
which overcomes the often arbitrary decomposition into "pair" and "embedding" contributions and thereby
allows one to compare different EAM-type potentials on the same scale.
More importantly this format offers a qualitative view on the energetics of pair interactions in a crystal 
(strictly, however, only for the perfect crystal structure and constant electron density).
The effective pair potentials for the MFS, Mendelev-II, and Chamati potentials are plotted in \fref{fig:closeup}(d).
Since the energy change of the stretched 2nn bond is much larger than for the other bonds and 
the energy contributions of the nn bonds and the compressed 2nn bond nearly cancel out each other, 
the overall energetics of the GSFE must be dominated by the stretched 2nn bond.
While being stretched, the energy contribution due to this bond is decaying and the contributions of the other bonds
become more important.
When external strains are applied, this transition takes place even earlier thereby opening the possibility that the stretched nn bonds
or the compressed 2nn bond have not yet passed through the minimum (which lies between the nn and 2nn distance for most potentials).
In such a case, the evolution of the total (GSFE) energy of the system would also pass through a (local) minimum.  

The analysis of the nn and 2nn distances in \fref{fig:closeup}(d) suggests that for an EAM-type potential to be robust against the formation of artificial \hkl{110} PFs,
one of the neighbor distances (nn or 2nn) should be significantly further apart from the minimum under zero-strain conditions than the other (as for the Chamati potential). 
At the same time, the "tail" of the effective potential should monotonically increase to zero at large distance, i.e., without additional terrace points
(e.g., Mendelev-II, Mendelev-IIext, and Marinica07).
This recipe would also explain our previous observation that the cutoff radius does apparently not play a major role for the formation of \hkl{110} PFs. 
The plots of the effective pair potentials for the EAM-type Fe potentials can be found in the [Supplemental Material].  
Although the effective pair potential is generally more suitable for a reliable and robust criterion than the decomposition into pair/embedding contributions,
it has to be noted that there are also cases where this recipe does not work:
Namely, in case of the FS potential for Cr where the minimum lies well-centered between the nn and 2nn distances but does not show the formation of \hkl{110} PFs. 

To overcome the lack of robustness of the previously suggested recipes, we introduce an empirical but straightforward criterion for the tendency to form \hkl{110} PFs.
An important feature of the potentials which exhibit a local GSFE minimum, 
which can be clearly observed in Figs.~\ref{fig:gsfe_Fe}--\ref{fig:gsfe_bcc}, is that the formation of a local GSFE minimum (or alternatively, the formation of a "hump") at applied strains is linked to the formation of a "shoulder" under unstrained conditions
(with "shoulder" we mean the region between two consecutive inflection points in the GSFE curves, 
where the curvature changes from positive to negative and back).
The Mendelev-II, Chiesa, M\"uller, and Bj\"orkas potentials for Fe and the Ackland potentials for Mo and W are prominent examples for this rather qualitative description, cf. Figs.~\ref{fig:gsfe_Fe} to \ref{fig:gsfe_bcc}.
In an attempt to quantify this visual distinction between "hump" and "shoulder", we calculate and compare the ratios 
\begin{eqnarray}
R = \frac{\gamma(x=\nicefrac{1}{6})}{\gamma(x=\nicefrac{1}{12})}\label{eq:R}
\end{eqnarray}
in \tref{tab:Delta_E} for selected material models for Fe and visualize them in \fref{fig:decompose_gsfe} for the EAM potentials by Mendelev \etal{} and Chamati \etal{} and the BOP by Mrovec \etal{}. 
When comparing the $R$ values for the different models, it becomes evident that a significant decrease of $R$ (more than 40\%) from unstrained to strained conditions is indicative for the formation of a local GSFE minimum. 
Moreover, $R < 2$ is a good indicator for a "shoulder" already under unstrained conditions, as, e.g., for the Bj\"orkas potential.
An $R$ value around or below unity is indicative for a "hump," as e.g. for the Mendelev-II and Chiesa potentials. 
This means, that for the future development of EAM potentials (and likewise for the pair-wise parts of MEAM, ADP and Tersoff potentials) 
it may be enough to ensure that $R$ is well above 3 under unstrained and strained conditions which does not 
involve the calculation of the complete GSFE surface but only at $x = 0$, 1/12, and 1/6.

Since we have not yet found a conclusive and generally applicable explanation for the formation of the local GSFE minimum under applied strains, 
we recommend including the determination of the strain-dependent \hkl(110) GSFE in the \hkl[1-10] direction (or only the $R$ value for reasons of efficiency) 
as a benchmark for newly developed potentials until the underlying reasons are clarified.
For many practical applications, it may even be sufficient if $R$ remains above 3 for equi-biaxial strains up to around 6\%--7\%.
Potentials, which show PFs only at such high strains, can be seen as effectively artifact-free in this respect; see, e.g., the Chamati potential for Fe, which gives a local GSFE minimum at 7.5\% equi-biaxial strain (cf. \tref{tab:overview}), 
but no PFs under practical conditions (cf. \fref{fig:cracktips_fcc-vs-110pf}), i.e., at crack tips. 
 
\begin{table}
\caption{Comparison of $R$ [cf. \eref{eq:R}] for different material models for Fe.
\label{tab:Delta_E}}
\begin{tabular}{cccc}
\hline
\hline
Model & Parameter set & $R(\varepsilon$=$0\%)$ & $R(\varepsilon_\text{EBS}$=$5\%)$\\
\hline
EAM & Mendelev-II & 2.9 & 0.4  \\
EAM & Chamati & 3.6 & 2.1 \\
EAM & Chiesa & 3.8 & 1.3 \\
MEAM & Lee2001 & 2.8 & 2.6 \\ 
ADP & Mishin & 3.6 & 3.1\\
Tersoff & Bj\"orkas & 1.6 & 1.4$^*$\\
BOP & Mrovec & 4.0 & 3.2 \\
DFT & USPP-PBE & 3.2 & 3.2 \\
\hline
\hline
\end{tabular}
\\
$^*$ both $\gamma(x=1/6)$ and $\gamma(x=1/12)$ are negative
\end{table}
 
\section{Summary}
\label{sec:summary}

In this study, we determined the strain dependence of the generalized stacking fault energy (GSFE) of \hkl{110} planes  
in bcc transition metals with different state-of-the-art material models for atomic-scale simulations:
EAM and MEAM potentials, ADPs, Tersoff potentials, BOPs, and DFT calculations. 
We showed that a number of EAM, MEAM, and Tersoff potentials predict an evolution of the strain-dependent \hkl{110} GSFE 
that exhibits a local minimum under applied uni- and equi-biaxial tensile strains resulting in the formation of planar faults (PFs), 
which are structurally very similar to the fcc structure. 
Examples for the formation of PFs in practice include simulations of cracks \cite{Lat03,Bor11,Vat11,Vatne2011,Moe14msmse,Ers12,Moe15,Motasem2016} 
and nanowires \cite{Yang2016,Zhang2012a,Skogsrud2015}.
Since a local GSFE minimum under applied external strains is not observed with more sophisticated material models, i.e., DFT and BOPs,
we conclude that the strain-dependent formation of PFs is an artifact of many classical potentials.
For this reason, previous discussions based on observations of PFs (or misclassified fcc formations) in atomistic simulations can become questionable and should better be re-evaluated carefully. 

We show that a local GSFE minimum is not formed for reasons of symmetry and 
that the inclusion of angular-dependent interaction terms is necessary but not sufficient for a material model to exhibit an artifact-free GSFE.
For purely distance-dependent potentials, a too short cutoff radius is excluded as possible reason for the formation of \hkl{110} PFs.
Instead, we find that the evolution of the effective pair potential for the next- and second-next-nearest neighbor distances can be used to qualitatively understand the formation of a local GSFE minimum.
Our attempts to develop a robust recipe to prevent this behavior were, however, not fully conclusive and we hope for future research in this direction.
Until such a recipe is found, we recommend including the determination of the strain-dependent \hkl(110) GSFE 
(or the $R$ value introduced in this paper) as a benchmark for newly developed potentials.

\section*{Acknowledgements}
J.J.M. gratefully acknowledges financial support of the Deutsche Forschungsgemeinschaft (DFG) via the research training group GRK 1896 "In Situ Microscopy with Electrons, X-rays and Scanning Probes."
J.J.M. thanks B. Meyer for the pseudopotentials for Fe, Nb, Mo, and W, T. Kl\"offel for invaluable comments on the DFT calculations as well as J. Gu\'enol\'e and W. G. N\"ohring for technical support.
T.H., R.D., and E.B. acknowledge financial support of the DFG through project C1 and C3 of the collaborative research centre SFB/TR 103 "From Atoms to Turbine Blades."
I.B., T.H., and J.N. acknowledge financial support of the DFG within the SFB 761 "Steel--ab initio."

\bibliographystyle{vancouver}
\bibliography{library}

\end{document}